\documentclass[twocolumn,showpacs,preprintnumbers,amsmath,amssymb]{revtex4}

\usepackage{graphicx}
\usepackage{dcolumn}
\usepackage{bm}

\begin{document}

\title{Semiclassical dynamics of atomic Bose-Einstein condensates}
\author{S. Choi and B. Sundaram}
\affiliation{Department of Physics, University of Massachusetts, Boston, MA 02125, USA}

\begin{abstract}
An atomic Bose-Einstein condensate (BEC) is often described as a
macroscopic object which can be approximated by a coherent state.
This, on the surface, would appear to indicate that its behavior should be
close to being classical.  In this paper, we clarify the extent of
how ``classical'' a BEC is by exploring the semiclassical equations for BECs under
the mean field Gaussian approximation. Such equations describe the dynamics of a condensate
in the classical limit in terms of the variables $\langle x \rangle$ and $\langle p \rangle$
as well as their respective variances.  We compare the semiclassical solution with the full quantum solution
based on the Gross-Pitaevskii Equation (GPE) and find that the interatomic interactions which generate nonlinearity make the system less ``classical.''  On the other hand,
many qualitative features are captured by the semiclassical
equations, and the equations to be solved are far less
computationally intensive than solving the GPE which make them ideal for
providing quick diagnostics, and for obtaining new intuitive insight.
\end{abstract}


\maketitle


An atomic Bose-Einstein condensate (BEC) is a macroscopic quantum object  composed of typically millions of bosonic atoms.
By now it is well-established  that a large class of phenomena involving atomic BECs can be
described accurately by the Gross-Pitaevskii Equation (GPE) for the condensate
mean field\cite{Review}. The incoherent component outside the mean field can, in most situations, be ignored as they become significant only at
higher temperatures closer to the critical temperature of BEC phase transition\cite{finiteT}.
The GPE, which is a nonlinear Schr\"{o}dinger Equation for the condensate wave function $\psi$ in the presence of an external potential $V_{ext}(x)$, is
\begin{equation}
i \hbar \frac{\partial \psi}{\partial t} = \left [
-\frac{\hbar^2}{2m} \nabla^2 + V_{ext}(x) + U_0 |\psi|^2 \right ] \psi  \label{GPE}
\end{equation}
where $U_0 = 4 \pi \hbar^2 a N_0/m$ with $a$ being
the $s$-wave scattering length; $N_0$ is the total number of
atoms in the condensate.

To derive the GPE, the quantum state of a BEC is assumed to well-approximated by the coherent
state\cite{Coherent}; the best  justification for this choice of quantum state is of course, the success of the GPE itself.
There are, in fact,  a number of unique characteristics  associated with the
coherent state, and the one that has gained most attention in
the context of BEC studies is that of the phase coherence  -- an indication that the many-body quantum system is
acting  effectively   as a single quantum object.
Another important characteristic of the  coherent state is that it is the
quantum state closest to the corresponding classical system.
A coherent state $| \alpha \rangle$ can  be shown to be identical to minimum uncertainty Gaussians where:
\begin{equation}
{\rm Re} [\alpha] = \langle x \rangle \left ( \frac{m \omega}{2
\hbar} \right )^{1/2} \;\;\;\; {\rm and} \;\;\;\; {\rm Im} [\alpha]
= \langle p \rangle \left ( \frac{1}{2 m \omega \hbar} \right
)^{1/2} , \nonumber
\end{equation}
and it was Schr\"{o}dinger himself who first showed\cite{Milonni} that Gaussian
wave packets form coherent states of a harmonic oscillator that
display classical motion. Further studies  in various contexts~\cite{Heller,habib,PSpapers} have confirmed that Gaussian dynamics  is indeed classical at all levels\cite{Heller,habib}. We note that the ``Gaussian dynamics''  in this context includes a
``squeezed coherent state''  i.e. Gaussian of variable width, and the state is assumed to remain a Gaussian throughout the evolution.

Despite the coherent state/Gaussian ansatz, the solution to the GPE does not necessarily
give classical dynamics. This apparent ``contradiction'' arises simply because the coherent state assumption
used to derive the GPE is made at the level of second quantization -- the ansatz turns the many-body quantum field operator into a {\it classical field} of  condensate wave function.  The resulting wave function in real space can, in principle, take any form depending on the boundary and external conditions.
In order for the  classical dynamics to emerge from the condensate wave function, a further ``coherent state'' assumption {\it within} the mean-field level of theory -- that the condensate wave function takes on a Gaussian profile -- is needed.  The fact that  a BEC is not strictly  a ``coherent state'' within  the mean-field regime is evident from examining the Q-function\cite{Qfunc} for the condensate at the level of GPE:
a Gaussian wave packet gives a circular Q-function  expected of a coherent state  while the Thomas-Fermi approximation for which most experiments are valid,  gives an elliptical Q-function indicative of its nonclassical nature\cite{ChoiSqueeze}.

The Gaussian profile approximation can, in fact, be a good
approximation for the stationary condensates with a small number of
atoms or with an attractive interatomic interaction. However, the
inherent nonlinearity of the system can destroy any Gaussian profile
approximation very quickly during the dynamical evolution. We note
that a Gaussian description was used previously in the context of
studying surface collective mode of a BEC with attractive
interactions\cite{PRiceExp}, where the results based on variational
methods\cite{variational1,variational2} were used.  In the early
works based on variational
principle\cite{variational1,variational2}, it was shown that many of
the features of a BEC can be captured using the Gaussian ansatz.
However, the semiclassical aspect and the validity of the Gaussian
dynamics compared to the full quantum dynamics was not fully
explored. Also the results were restricted to the harmonic trapping
potential, which doesn't demonstrate a more general picture of
deviation from the GPE.

In this paper,  we ask the question ``Just how classical is a BEC
and under what conditions?'' We explore the dynamics of the atomic
BECs in the semiclassical limit and compare with the full quantum
solution described by the GPE. The semiclassical limit is obtained
by assuming that a BEC maintains a Gaussian profile (of variable
mean and variance) throughout the evolution and deriving the
corresponding effective Gaussian semiclassical equations. Strictly
speaking, only the ground state for a harmonic trap with $U_0 = 0$
in the GPE of Eq. (\ref{GPE}) is precisely Gaussian; for $U_0 \neq
0$, a Gaussian profile  provides an  idealized classical limit that
acts as a ``baseline'' for comparison with the full quantum result.
The motivation for this study comes not just from addressing
fundamental questions on the quantum-classical nature of a BEC but
also in the context of advancing quantum engineering and control
methods\cite{Choi1} via simpler equations that relate relevant
observables. The semiclassical equations can provide us a means to
understand complex quantum behavior in terms of the more familiar
and intuitive classical picture, providing a new insight into how to
manipulate atomic BECs.

%

 The usual derivation of Dirac's
time-dependent variational principle states that  an action of
the form $\Gamma =\int dt \langle \Psi, t |i \hbar
(\partial/\partial t) - \hat{H}|\Psi, t \rangle$ $\delta \Gamma = 0$
results in the Schr\"{o}dinger Equation. The solution is obtained by
restricting $|\Psi, t \rangle$ to a subspace of the full Hilbert
space.  For deriving the semiclassical Gaussian  dynamics, the subspace is assumed to be that of a ``squeezed state''
wave function of the form\cite{PSpapers}:
\begin{equation}
\psi(x, t) = \frac{1}{(2 \pi \rho^2)^{1/4}} \exp \left \{ -
\frac{\alpha}{4  \rho^2}(x - \langle x \rangle)^2  + i p(x - \langle
x \rangle) \right \} \label{state}
\end{equation}
where, with the definition $\Delta A  \equiv A- \langle A
\rangle$, $\alpha = 1 - i  \langle \Delta x \Delta p + \Delta p \Delta x
\rangle$ and  $\rho$ is defined such that
\begin{equation}
\langle \Delta x^{2m} \rangle
= \frac{\rho^{2m}(2m!)}{2^{m}m!}
\end{equation}
with  $m = 1$ corresponding to the position variance, $\langle \Delta x^{2} \rangle  = \rho^2$.
The use of a squeezed state ansatz is reminiscent of the
time-dependent Hartree-Fock-Bogoliubov (TDHFB) theory\cite{Chernyak}.
The major difference is that with TDHFB  the squeezed state ansatz is applied for both the condensate and non-condensate atoms, as it is a finite
temperature theory. Further, the mean field solution used is the precise ground state for the
system, not a Gaussian approximation i.e. it has nothing to do with the
classical dynamics. TDHFB is also a very difficult problem to solve.
Here, we are interested in
the semiclassical dynamics at the mean field level, i.e.
where all the atoms are assumed to be in the condensate.

With such an ansatz, the dynamics of the system characterized by a
Hamiltonian $H = p^2/2 + V(x)$ (here we set the mass to unity) can be represented as an extended
classical gradient system for the average and fluctuation variables.
The dynamical equations for the coordinate variable $\langle x \rangle$
and its standard deviation, $\rho$ are\cite{PSpapers}
\begin{eqnarray}
\frac{d^2 \langle x \rangle }{dt^2}  &  = & - \sum_{m = 0}^{\infty}
\frac{\rho^{2m}}{m! 2^{m}} V^{(2m+1)}(\langle x \rangle)   \label{dx2} \\
\frac{d^2 \rho}{dt^2}  & = & \frac{\hbar^2}{4 \rho^3} -  \sum_{m =
1}^{\infty} \frac{\rho^{2m-1}}{(m-1)! 2^{m-1}} V^{(2m)}(\langle x
\rangle)  \label{drho2}
\end{eqnarray}
where $V^{(n)}(\langle x \rangle) \equiv \left.
\frac{\partial^{n}}{\partial x^{n}} V(x) \right |_{x= \langle x
\rangle}$.

By integrating the above equations once with respect to time, the dynamics of the respective conjugate variables $\langle p \rangle$ and $\Pi$
are obtained:
\begin{equation}
\langle p \rangle  =  \frac{d \langle x \rangle }{dt} \;\;\;\; {\rm
and} \;\;\;\;  \Pi   = \frac{d \rho}{dt},  \label{firstderiv}
\end{equation}
where
\begin{equation}
\Pi \equiv \frac{\langle \Delta x \Delta p + \Delta p \Delta x
\rangle}{2 \rho} .  \label{Pi}
\end{equation}
From the uncertainty relation
 \begin{equation}
\langle \Delta x^{2} \rangle \langle \Delta p^{2} \rangle \geq
\frac{\hbar^2}{4}  + \frac{\langle \Delta x \Delta p + \Delta p
\Delta x \rangle^2}{4},
\end{equation}
the momentum variance can be written in terms of $\rho$ and $\Pi$ as
\begin{equation}
\langle \Delta p^{2} \rangle  = \frac{\hbar^2}{4\rho^2} + \Pi^2 ,
\end{equation}
 implying that
whenever $\Pi = 0$, we have the minimum uncertainty state. We note
that  Eqs. (\ref{dx2} - \ref{Pi}) are the most general equations of
motion for $ \langle x \rangle$, $\langle p \rangle$, $\rho$ and
$\Pi$ for a squeezed coherent state in an arbitrary potential
$V(x)$.  We use this set of self-contained equations to model our
case. It is noted that these equations may also be derived
alternatively using an operator expansion method\cite{SM}. A point
of interest is  how to include
 the interatomic interaction in the equations of motion.

As shown in Eq. (\ref{GPE}), the GPE for the condensate wave
function is a nonlinear Schr\"{o}dinger equation with an additional
mean field potential term $U_0  |\psi(x, t)|^2$. We therefore model
nonlinearity by extending our potential to introduce the
``internal'' mean field potential experienced by an atom in the
condensate i.e. $V(x,t) = V_{\rm ext}(x, t) +  V_{\rm int}(x, t)$
where $ V_{\rm ext}(x, t)$ is the usual externally imposed potential
(such as the harmonic trap) and $V_{\rm int}(x, t)$ is the internal
``mean field'' potential due to the interatomic interactions
experienced by the condensate atoms:
\begin{eqnarray}
V_{\rm int}(x, t) & \equiv  &  U_0 |\psi(x, t)|^2 =  \frac{U_0 e^{- \frac{(x - \langle x \rangle)^2  }{2 \rho^2} }}{\sqrt{2 \pi \rho^2}} ,
\end{eqnarray}
where we have assumed that the condensate wave function is given by the squeezed state
 Eq. (\ref{state}).
 Since the semiclassical equations are essentially equations governing $\langle x
\rangle$ and $\rho$ and $V_{\rm int}(x, t)$ is itself a function of
the time-dependent variables $\langle x \rangle$ and $\rho$, it
should provides the necessary nonlinear effect consistent within
this framework.  We shall check this by comparing with the equations
derived independently in Ref. \cite{variational2} where the
Lagrangian itself contained the nonlinear interaction term from the
outset.

For the purposes of deriving the semiclassical equations for atomic BECs, we shall consider
a most general (practical) form for the external potential:
\begin{equation}
V_{\rm ext}(x,t)  =  \frac{1}{2} \omega^2 x^2  -\alpha  [ \cos (kx -  \Omega t) - 1] + \beta x
   \label{ExtV}
\end{equation}
where $\omega$ provides the frequency of the harmonic
trap, $\alpha$ is the amplitude of an optical lattice, $\beta$ is the gradient of the tilt of the optical
lattice, and $\Omega$ allows the possibility of a travelling lattice at
speed $\Omega/k$. By setting $\alpha = \beta = \Omega = 0$ and $\omega \neq 0$
one is considering the case of a harmonically trapped condensate,
while setting $\alpha \neq 0$ and $\omega = 0$ would imply the
presence of an optical lattice only. It is noted that all of these
parameters $\alpha$, $\beta$, $k$, $\Omega$ and $\omega$ can, in
general, be time-dependent quantities. The form of the optical lattice
potential was chosen so that at time $t = 0$ it approximates the standard harmonic
potential for a condensate initially at mean position $\langle x
\rangle = 0$.

It was found that the infinite series of Eqs. (\ref{dx2}) and (\ref{drho2}) involving the $V_{\rm ext.}(x,t)$ part can be obtained
relatively straightforwardly, while that involving $V_{\rm int}(x, t)$ was less straightforward, and required the use of the definition of the Hermite polynomials $H_n(x)$,
\begin{equation}
H_{n} (x)  =  (-1)^n e^{x^2} \frac{d^{n}}{dx^{n}} e^{-x^2},
\end{equation}
the identity
\begin{equation}
H_{2n+1} (x)  =  (-1)^n 2^{2n + 1} n! x L^{(1/2)}_{n}(x^2),
\end{equation}
where  $L^{(\alpha)}_{n}(x)$ are the generalized Laguerre polynomials, and
the generating function for the Laguerre
polynomials
\begin{equation}
\frac{e^{-xt/(1-t)}}{(1 - t)^{\alpha + 1}} =    \sum_{n = 0}^{\infty}
L^{(\alpha)}_{n}(x) t^n .
\end{equation}
After some work, the semiclassical equations that describe a BEC in the general
external potential and in the presence of a mean field is finally obtained as:
\begin{eqnarray}
\frac{d^2  \langle x \rangle }{dt^2} &  = & -\alpha k \sin(k  \langle x \rangle - \Omega t)  e^{-\rho^{2}k^{2}/2}  - \omega^2  \langle x \rangle  - \beta  \label{dp}\\
\frac{d^2 \rho}{dt^2}   & = &  - \left [ \alpha k^2 \cos(k \langle x \rangle - \Omega t)  e^{-\rho^{2}k^{2}/2} + \omega^2 \right ] \rho  \nonumber \\
                  & & + \frac{\hbar^2}{4 \rho^3} + \frac{ U_0 }{4\sqrt{\pi} \rho^3 }  . \label{dPi}
\end{eqnarray}
These equations in  $\langle x \rangle$ and $\rho$  can be very easily solved numerically  compared to the GPE which requires a dense
 grid in both position and time for accuracy. In particular, the
 equations are essentially identical to those obtained in Ref.
 \cite{variational2}. This proves the validity of the assumption
 that at least for the Gaussian anstz, the effect of nonlinearity
 can be modeled exactly by assuming
 that the interatomic interaction effectively
  results in a  modified potential. This insight can help in analyzing equations in other
  contexts.

The variables in the semiclassical equations yield
 an extended potential system where the
fluctuation and average variables are treated on equal footing.
The
extended Hamiltonian $H_{ext.} \equiv \langle H \rangle$ in our case is given by
$H_{ext.} = H_{ x } + H_{\rho} + H_{ x  \rho}$ where
\begin{eqnarray}
H_{ x } & =  & \frac{\langle p \rangle^2}{2}
 + \beta \langle x \rangle + \frac{1}{2}
\omega^2 \langle x \rangle ^2    \\
H_{ x  \rho}  & =  &   - \alpha \cos(k \langle x \rangle -
\Omega t) e^{-\rho^{2}k^{2}/2}  \\
H_{\rho} & =  &  \frac{\Pi^2}{2}  + \frac{1}{2}\omega^2 \rho^2
+ \frac{\hbar^2}{8 \rho^2} + \frac{U_0}{8\sqrt{\pi}\rho^2}  . \label{Hrho}
\end{eqnarray}
The form of the extended Hamiltonian indicates effectively an infinite
barrier potential $H_{ext.} \rightarrow \infty$ as $\rho \rightarrow 0$ i.e. the ``quantum fluctuations'' can never go to
zero except in the limit $\hbar \rightarrow 0$  and $U_0 \rightarrow 0$. The last two
terms of Eqs. (\ref{dPi}) and (\ref{Hrho}) indicate that the effect of having a mean
field interatomic interaction can be thought of as effectively modifying the
Plank constant $\hbar \rightarrow \hbar_{eff.}$ such that:
\begin{eqnarray}
\hbar_{eff.} = \sqrt{\hbar^2 + \frac{U_0}{\sqrt{\pi}}} .
\end{eqnarray}
This formally supports the observation that nonlinear interatomic
interactions (having $U_0 >0$) push the system deeper into the
``quantum'' regime.



Several interesting observations can be made simply by examining
Eqs. (\ref{dp}) and (\ref{dPi}) a bit more closely. For instance,
 it is clear
how to directly set the accelerations,  $d^{2}\langle x
\rangle/dt^{2}$ and $d^{2}\rho/dt^{2}$ by choosing various initial
parameters $\alpha$, $k$, $\Omega$, $\omega$, $\beta$ and $U_0$.  This
can be useful in designing quantum control strategies to manipulate a wave packet's acceleration.
Such direct and intuitive information is
not readily available from  the GPE.
In particular, the entire right hand side of  Eq.
(\ref{dPi}) can be made to vanish i.e. one can figure out how to
stop the variance of a wave packet  from accelerating.

This is significant since,
if one starts off with a stationary state i.e. with a zero initial
rate of change $d \rho/dt = 0$,  the width of the wave packet
throughout the evolution is ``frozen'' at the initial value under such conditions. For a Gaussian this
is tantamount to having its overall shape unaltered throughout its evolution.
Although such a behavior is reminiscent of a soliton solution\cite{Soliton},
they are better classified as solitary
waves\cite{SolitaryWave} as it has less to do with the nonlinearity of the system and more to do
with being a stationary state of the extended Hamiltonian. Another way to look at this is to consider the
gradient of the extended potential with respect to the position
variance. When  one has
\begin{equation}
\frac{d H_{x \rho}}{d \rho} + \frac{d H_{\rho}}{d \rho}  = 0 ,
\end{equation}
$\rho$  continues on with its initial value without acceleration.

\begin{figure}
\begin{center}
\centerline{\includegraphics[height=7cm]{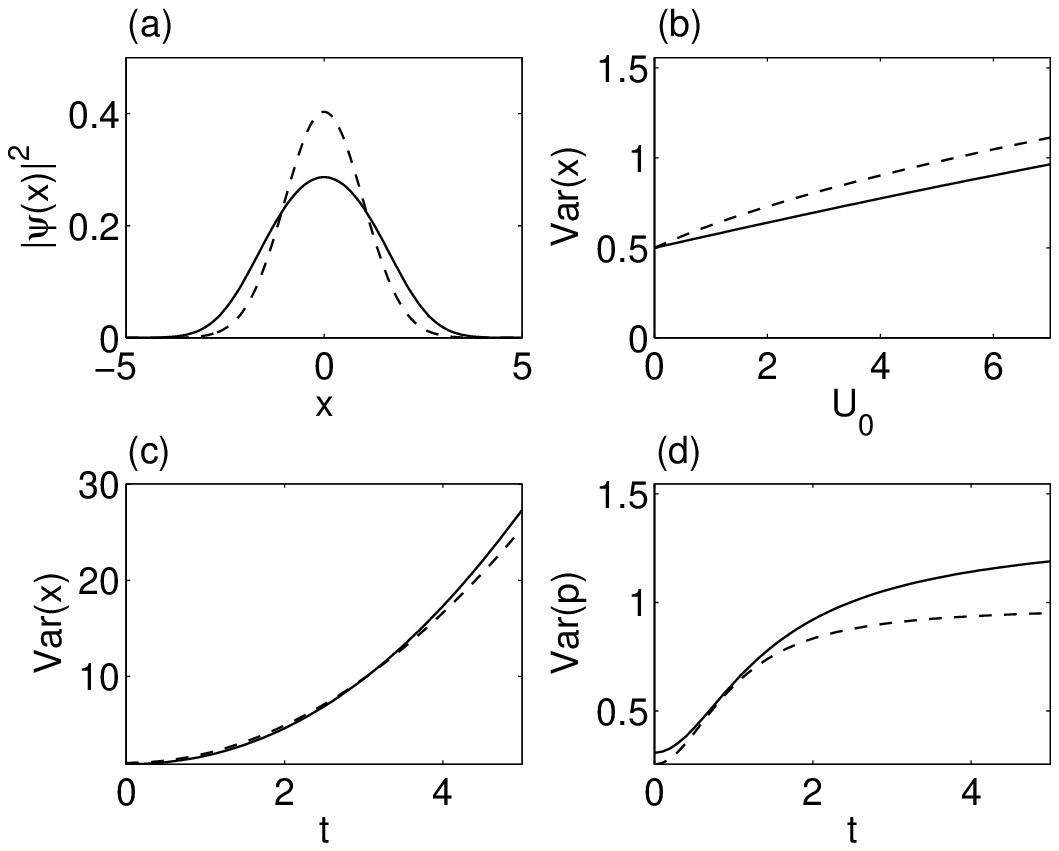}} \caption{Solid line represents
the GPE solution and dashed line represents the Gaussian semiclassical prediction.  (a)
Position probability density $|\psi(x)|^2$ for the ``stationary state'' corresponding to $U_0 = 5$.
(b) the width of the stationary state wave packet represented by the position variance for various values of $U_0$.
(c) evolution of the position variance of the $U_0 = 5$ wave packet after being released from the trap. (d)
 evolution of the momentum variance of the $U_0 = 5$ wave packet after being released from the trap. Time is in units of harmonic oscillator period throughout this paper.} \label{ballexp}
\end{center}
\vspace{-0.75cm}
\end{figure}

For the case
of a harmonically trapped condensate ($\alpha = \beta = 0$, $\omega > 0$)
the right hand side of Eq. (\ref{dPi}) vanishes when the initial width of the wave packet is such that
\begin{equation}
\rho = \sqrt{\frac{h_{eff.}}{2 \omega}}
\end{equation}
i.e. it gives the expected width
of the stationary state solution for a given value of $U_0$.  In Fig. \ref{ballexp} the prediction of the
semiclassical equation for the stationary state solution is compared
with those obtained using the full quantum ground state solution of
the GPE\cite{Pit}. Figure \ref{ballexp}(a)compares the shape of the
Gaussian solitary wave solution for $U_0= 5$ with the ground state
of GPE for the same value of nonlinearity, while  Fig. \ref{ballexp}(b) compares the width of the stationary state wave function as a function of
$U_0$.   We have checked using the numerical ground state solution to
GPE with a harmonic trap\cite{Pit} that for $U_0$ up to around 9, the profile of the
condensate can be well-approximated by a Gaussian before going over to
the inverted parabolic form of the Thomas-Fermi regime.


 A common experimentally relevant situation is when the
harmonic trap is turned off in the absence of an optical lattice,
($\alpha = \beta = 0$,  $\omega = 0$,  $U_0 > 0$). It is easy to see from the
right hand side of  Eq. (\ref{dPi}) being positive that the position
variance $\rho$ will start to increase without bound when the trap is suddenly turned off. In addition, as $\rho \rightarrow \infty$,  $d^2
\rho/dt^2 \equiv d \Pi/dt \rightarrow 0$ due to the last two terms of Eq.
(\ref{dPi}) i.e. $\Pi$ becomes a constant, and consequently
\begin{equation}
\langle \Delta p^{2} \rangle
\equiv \frac{\hbar^2}{4\rho^2} + \Pi^2 \rightarrow  \Pi^2
\end{equation}
 i.e.  it is expected that the momentum
variance reaches a constant value after a while. Although this fact about the momentum variance
is not at all obvious from the GPE or even from the theory of Fourier transforms -- one would suspect the momentum variance tends towards zero as the position variance increases without bound -- Fig. \ref{ballexp}(c) and (d)  indicate that the quantum simulation using GPE
shows a good  qualitative agreement with the semiclassical prediction.

What is also notable is that, if one wants to hold the wave packet together
in the absence of any confining
potential ($\alpha = \beta = 0$, $\omega = 0$), one needs $U_0 = -
\hbar^2\sqrt{\pi} < 0$  to get the right hand side of Eq. (\ref{dPi}) to vanish  i.e. an
attractive interaction, as expected. This also
presents the limit as to how far negative $U_0$ can be for a
harmonically trapped BEC with an attractive interaction. This is
because, with a harmonic potential present ($\alpha = \beta = 0$, $\omega >
0$),  $U_0 = - \hbar^2 \sqrt{\pi}$ in Eq. (\ref{dPi}) eliminates the
infinite barrier potential so that $\rho$ can take on zero value
after some time. This then makes $\langle \Delta p^{2} \rangle
\equiv
\frac{\hbar^2}{4 \rho} + \Pi^2$ to diverge, making the system
unstable -- this limit on $U_0$  shows that, according to a
semiclassical argument, a given harmonically trapped BEC with a
negative scattering length can support up to
\begin{equation}
N =\frac{1}{4\sqrt{\pi}|a|} \approx \frac{0.141}{|a|}
\end{equation}
 atoms where the interatomic scattering length $a$ is in harmonic oscillator lengths $\sqrt{\hbar/m\omega}$.  Ignoring additional considerations such as the dimensionality of the condensate this is of comparable order of magnitude to the
prediction made using the full quantum theory\cite{Stoof} for the upper limit in atom number,
$N = 0.573 /|a|$. Again the semiclassical equations can be a useful tool in finding
order-of-magnitude predictions based on intuitive argument.


\begin{figure}
\begin{center}
\centerline{\includegraphics[height=7cm]{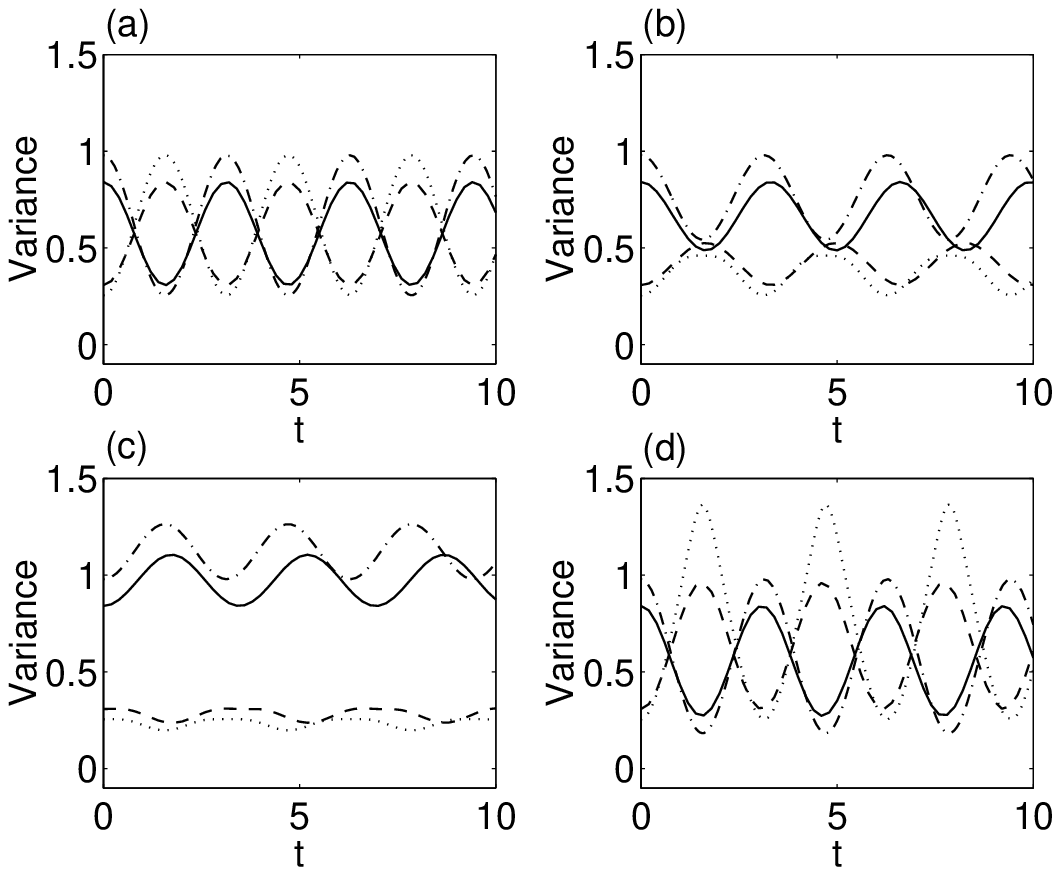}}
\caption{Time evolution of $\langle \Delta x^2 \rangle$ (Solid
line for the GPE simulation and dash-dot lines for the semiclassical prediction), and
$\langle \Delta p^2 \rangle$ (dashed line for the GPE simulation and dotted line for the semiclassical prediction) expected when a BEC with the initial nonlinearity of $U_0 = 5$ is tuned to new nonlinearity $U_{new}$ at time $t = 0$.  (a) $U_{new} = 0$,
(b) $U_{new} = 2$, (c) $U_{new} = 7$, and (d) $U_{new} = -0.5$  } \label{Variance1}
\end{center}
\vspace{-0.75cm}
\end{figure}

In a previous work\cite{Choi2},
the case where the interaction strength $U_0$ of a harmonically trapped atomic BEC is
suddenly tuned to zero was studied using the full quantum theory. In that work it was noted that a BEC starts to undergo oscillatory motion which correspond to the rotation of the error contour.
We investigate here a similar effect using the semiclassical equations: starting with the stationary state with nonlinearity $U_0 > 0$, the nonlinearity is suddenly switched to  $U_{new}$ at time $t=0$ and the subsequent evolution of the position and momentum variances is observed. We consider four possible cases: (a) $U_{new} = 0$, (b) $0 < U_{new} < U_0$, (c)  $U_{new} > U_0$  and
(d) $- \hbar^2 \sqrt{\pi} < U_{new} < 0$  and compare  the quantum calculations with the semiclassical prediction.
We show in Fig.~\ref{Variance1} the evolution of the correlations
$\langle \Delta x^2 \rangle$,  and $\langle \Delta p^2 \rangle$
of a trapped BEC  with initial $U_0 = 5$ when the nonlinearity is  tuned at time $t = 0$ to $U_{new} = 0$, $2$, $7$,
and $-0.5$.  It is noted that for $U_{new} = 0$ shown in Fig.~\ref{Variance1}(a)  the
evolution of both position and momentum variances vary sinusoidally with equal
amplitudes, consistent with the previous observation on the rotation of the error contour,
 while for $U_{new} \neq 0$, the oscillation amplitudes are
uneven, showing the effect of squeezing.

This result clearly indicates that
the amount of quantum squeezing can be controlled by controlling the magnitude of
$U_{new}$ relative to $U_0$.  From further simulations with different values of $U_{new}$, it was found that, in general, with a decreasing $U_{new} < U_0$ such as Fig.~\ref{Variance1}(b) the {\it minimum} value that the position variance takes during the oscillatory dynamics
{\it decreases}, indicating position squeezing while, with
an increasing $U_{new} > U_0$ such as Fig.~\ref{Variance1}(c), the {\it maximum} value that
the position variance takes {\it increases} indicating momentum squeezing. This shows that the general effect of positive nonlinearity is to induce momentum squeezing.
 The case with a
negative $U_0$ such as Fig.~\ref{Variance1}(d) showed that the momentum variance quickly
diverges as $U_0$ becomes more negative.

All these behaviors may be understood from the structure of the
extended Hamiltonian $H_{ext}$ and the changes in the infinite barrier potential term in Eq. (\ref{dPi}).
It should be noted that the semiclassical predictions differ only in the amplitude for
the $U_{new} = 0$ case while for $U_{new} \neq 0$, there is also a phase difference so that over
a long time, the semiclassical prediction gets worse. This feature due to the nonlinear effect agrees with the result observed in a
different context of driven nonlinear systems\cite{SM}, and indicates that the semiclassical dynamics are better for predicting
short time evolutions.


Finally we consider the evolution of variances in an optical
lattice instead of a harmonic trap. A salient aspect from the semiclassical equations is that, $\langle x
\rangle$  and $\rho$ are coupled in the
presence of an optical lattice, unlike the case of a simple harmonic trap ($\alpha = \beta = 0$) for which $\langle x
\rangle$  and $\rho$ are completely decoupled.  Also, due to the $ e^{-\rho^{2}k^{2}/2}$  factor, the effect
of the optical lattice is  significantly diminished when the wave
packet is extended over the lattice i.e. when $\rho^2 k^2 \gg 1$ or $\rho
\gg \lambda_{OL}$ where $\lambda_{OL}$ is the wavelength of the
optical lattice. Likewise, when $\rho$ is small, the effect of the optical
lattice becomes more pronounced.

\begin{figure}
\begin{center}
\centerline{\includegraphics[height=7cm]{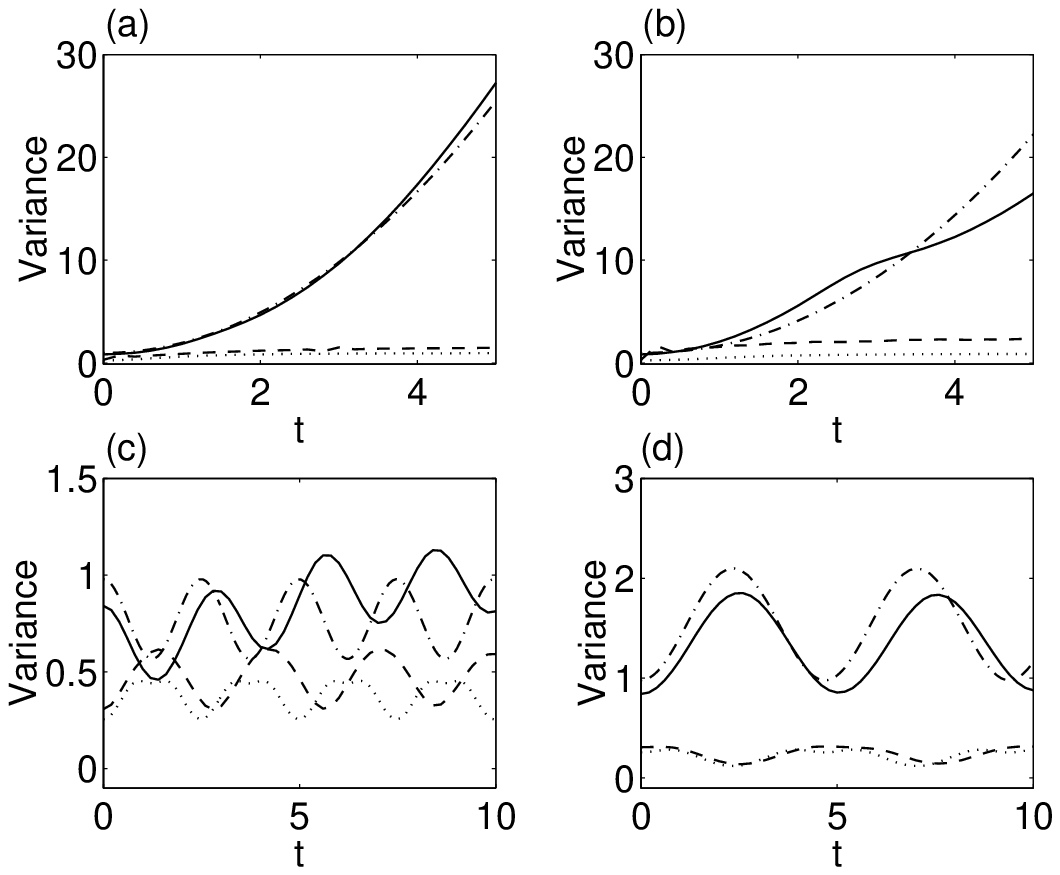}} \caption{Time evolution of  $\langle \Delta x^2 \rangle$ (Solid
line for the GPE simulation and dash-dot lines for the semiclassical prediction), and
$\langle \Delta p^2 \rangle$ (Dashed line for the GPE simulation and dotted line for the semiclassical prediction) expected when a BEC with the initial nonlinearity $U_0 = 5$ and position variance $\rho^2$ is placed  in an  optical lattice of different wavelengths, $\lambda_{OL}$.  (a) $\lambda_{OL} = \rho$,
(b) $\lambda_{OL} = 2\rho$, (c) $\lambda_{OL} = 10\rho$, and (d) $\lambda_{OL} = 20\rho$.} \label{Variance2}
\end{center}
\vspace{-0.75cm}
\end{figure}

\begin{figure}
\begin{center}
\centerline{\includegraphics[height=7cm]{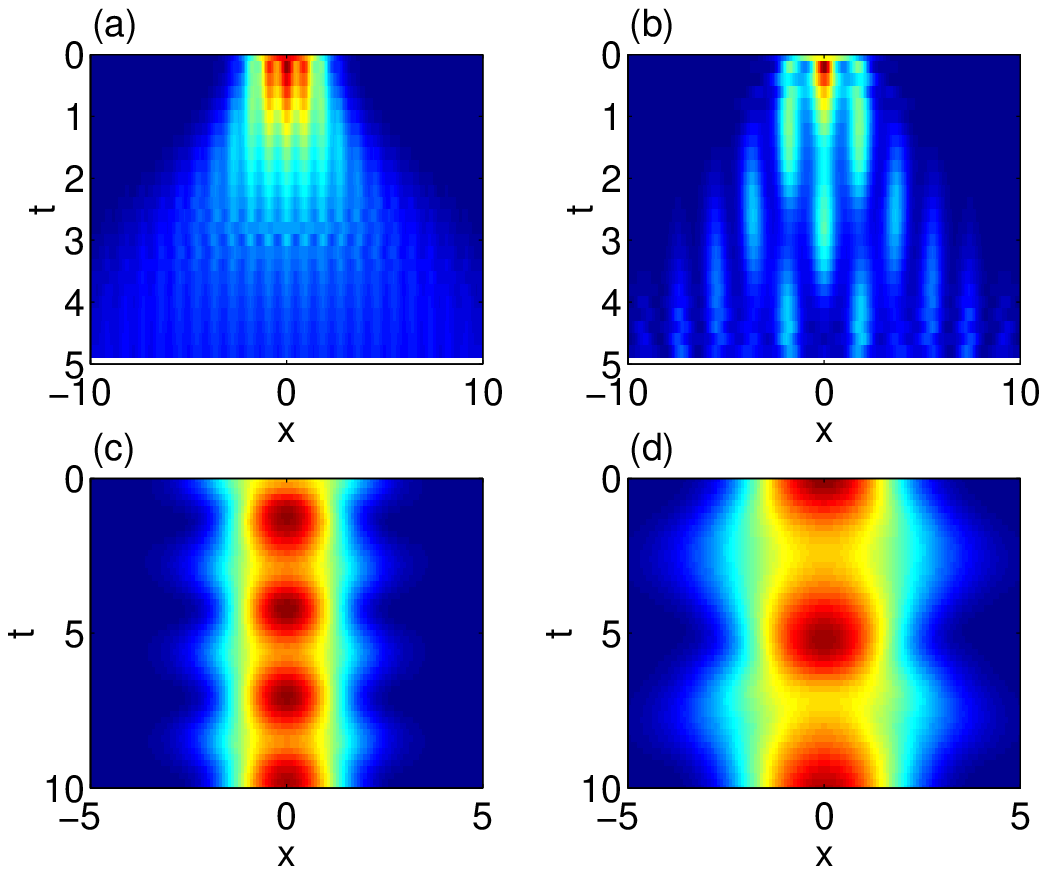}} \caption{ (Color online) Spatio-temporal plot of the probability density $|\psi(x,t)|^2$ obtained from the GPE when a BEC with the initial nonlinearity $U_0 = 5$ and position variance $\rho^2$ is placed in an optical lattice with wavelength (a) $\lambda_{OL} = \rho$,
(b) $\lambda_{OL} = 2\rho$, (c) $\lambda_{OL} = 10\rho$, and (d) $\lambda_{OL} = 20\rho$.  This shows that especially for $\lambda_{OL} = \rho$ and
$2\rho$ the Gaussian approximation is not valid.} \label{Profile}
\end{center}
\vspace{-0.75cm}
\end{figure}

Figure \ref{Variance2} shows what happens when one places a condensate corresponding to $U_0 = 5$ and position variance $\rho^2$  in an optical potential of varying wavelengths. It was found that for wavelengths $\lambda_{OL} < \rho$ it is as if the optical lattice is not present, and shows ballistic expansion-like behavior. In such cases, the GPE simulation agreed very well with the semiclassical prediction. However, as Fig. \ref{Profile}(a) shows, although the semiclassical equation captures
the mean and variance it does not capture the detailed structure formed in the wave packet. For  $\lambda_{OL} = 2\rho$, we see in Fig. \ref{Variance2}(b)    that the semiclassical prediction starts to deviate from the calculation of the full GPE solution; this is understandable from Fig. \ref{Profile}(b) where it shows the breakup of the wave function such that the Gaussian approximation cannot be valid (In fact, the Gaussian approximation is not valid for the $\lambda_{OL} = \rho$ case as well). For  $\lambda_{OL} = 10\rho$ and  $\lambda_{OL} = 20\rho$ shown in Fig.~\ref{Variance2}(c) and (d)  the wavelength is sufficiently large that the wave packet gets ``trapped'' in one of the wells and since the initial state is not an eigenstate of the optical lattice the evolution in time was again
oscillatory  similar to that shown in Fig. \ref{Variance1}.
We found in the case of $\lambda_{OL} \gg \rho$  that the smaller the value of the amplitude of the optical lattice
 $\alpha$ the greater is the position variance, indicating momentum squeezing; conversely, with a larger $\alpha$
one has increased momentum variance and position squeezing. A
possible related phenomenon is the Superfluid-Mott insulator
transition\cite{SFMott} (or a Josephson Junction as a two-site ``lattice'', with
the Rabi, Josephson, and Fock regimes in the order of increasing
 ratio of nonlinearity to tunnelling rate\cite{Choi3}.)  To verify whether the Josephson Junction physics can be captured using semiclassical dynamics,
 a new  set of equations  that assumes a Gaussian superposition  state needs to be derived, and will be dealt with in a future work.

 The example with the optical lattice
 shows that there is rich physics to be explored, and that one needs to exercise caution in applying the semiclassical analysis to make detailed experimental predictions. The results show clearly that the quantitative differences between the semiclassical and the quantum result originate from the fact that the semiclassical limit is more or less a ``coarse grained'' representation of the full quantum dynamics. This is unavoidable since, fundamentally,  classical mechanics deal with point particles in phase space whereas   quantum mechanics  deal with  wave functions. Indeed, the semiclassical approach such as the one discussed in this paper has been  originally developed as one of the many  attempts to bridge the gulf between the classical and quantum physics.


In summary, we have derived a set of equations for atomic BECs
which describe the condensate dynamics in the semiclassical limit.  The semiclassical equations have been derived by applying the Gaussian squeezed coherent state ansatz for the condensate wave function. From the comparison of the semiclassical dynamics with the expected full quantum dynamics,
it was found that the interatomic interactions tend to push the condensate deeper into the quantum regime.
The advantage of these equations was that it was easy to get a
more intuitive picture of the behavior of the system under
consideration.  Intuitive understanding could be obtained either from the classical Hamiltonian energy landscape argument or from simply examining the equations carefully and considering
the variables in various limits. Such an analysis is not possible with the GPE.
Despite the additional advantages such as the very high
numerical efficiency, if one wants to use the semiclassical equations as a predictive guide
to the full quantum dynamics, one has to make
sure that the Gaussian approximation is likely to be valid  throughout the evolution.
With this caveat, the semiclassical equations provide a quick diagnostics and  a verifiable ``baseline'' with which one can gauge the classical or quantum nature of
a condensate dynamics  under different situations.  \\


SC wishes to thank R. Reddy for discussions.


\end{document}